\documentclass{article}
\usepackage{icrctc07}

\newcommand{\gray}[1]{$\gamma$-ray{#1}}

\newcommand{\pubjournal}[6] {#1, #2 {\bf #3}, #4 (#5).}

\newcommand{\icrc}{{\it Int. Cosmic Ray Conf.}}

\newcommand{\aap}{{\it Astron. Astrophys.}}
\newcommand{\apj}{{\it ApJ}}
\newcommand{\sci}{{\it Science}}
\newcommand{\nat}{{\it Nature}}
\newcommand{\adv}{{\it Adv. Space Res.}}

\title{The Diffuse Galactic Gamma-Ray Emission Model for GLAST LAT}
\shorttitle{GLAST LAT Diffuse Emission Model}

\authors{
T. A. Porter$^{1}$, 
S. W. Digel$^{2,5}$, 
I. A. Grenier$^{3}$, 
I. V. Moskalenko$^{4,5}$, 
and A. W. Strong$^{6}$\\
for the GLAST LAT Collaboration}
\shortauthors{Porter et al.}
\afiliations{
$^1$Santa Cruz Institute for Particle Physics, University of California, Santa Cruz, CA 95064, U.S.A.\\
$^2$Stanford Linear Accelerator Centre, 2575 Sand Hill Road, Menlo Park, CA 94025, U.S.A.\\
$^3$AIM, Service d'Astrophysique, CEA Saclay 91191 Gif Sur Yvette, France\\
$^4$Hansen Experimental Physics Laboratory, Stanford University, Stanford, CA 94305, U.S.A.\\
$^5$Kavli Institute for Particle Astrophysics and Cosmology, Stanford University, CA 94309, U.S.A.\\
$^6$Max Planck Institut f\"{u}r extraterrestrische Physik, Postfach 1312, D-85741, Garching, Germany}
\email{tporter@scipp.ucsc.edu}

\abstract{Diffuse emission from the Milky Way dominates the \gray{} sky. 
About 80\% of the high-energy luminosity of the Milky Way comes from 
processes in the interstellar medium. 
The Galactic diffuse emission traces interactions of energetic particles, 
primarily protons and electrons, with the interstellar gas and radiation 
field, thus delivering information about cosmic-ray spectra and 
interstellar mass in distant 
locations. 
Additionally, the Galactic diffuse emission is the celestial foreground for 
the study of \gray{} point sources and the extragalactic diffuse \gray{}
emission. 
We report on the latest developments in the modelling of the Galactic 
diffuse emission, which will be used for the 
Gamma Ray Large Area Space Telescope (GLAST) investigations.}

\begin{document}
\maketitle

\section{Introduction}

Diffuse Galactic emission (DGE) dominates the \gray{} sky with more than 
80\% of the total luminosity coming from processes in the interstellar 
medium (ISM). 
The DGE is a tracer of energetic interactions of cosmic 
ray (CR) particles
in the ISM, and is produced by inverse Compton scattering (IC), 
bremsstrahlung, and $\pi^0$-decay.
It delivers information about spectra and intensities of CR species at distant
locations and allows the study of CR acceleration in sources as well as
propagation in the ISM.
Gamma rays can be used to trace the interstellar gas independently of 
other astronomical methods, e.g., the relation of the molecular H$_2$ gas
to CO \cite{Strong2004} and hydrogen overlooked by 
other methods \cite{Grenier2005}.
Additionally, 
the DGE is the bright ``background'' 
against which \gray{} point sources are detected and its accurate 
determination is important for localisation of such sources and their 
spectra, especially at low Galactic latitudes.
Furthermore, the DGE acts as a ``foreground'' for 
any extragalactic signal that we seek to recover.

%The majority of the DGE 
%is produced in energetic CR interactions, mainly protons 
%and electrons, with the interstellar gas and radiation field. 
%It carries
%the information about particle spectra in distant regions of the Galaxy.
Calculation of the DGE requires a model of 
CR propagation. 
Such models are based on the theory of particle transport and interactions 
in the ISM as well 
as many kinds of data provided by different experiments in Astrophysics and
Particle and Nuclear Physics.
Such data include: 
secondary particle and isotopic production cross sections, total 
interaction nuclear cross sections and lifetimes of radioactive species, 
gas mass calibrations and gas distributions in the Galaxy (H$_2$, H~I, H~II), 
interstellar radiation field (ISRF), CR source distribution and particle
spectra at the sources, and the Galactic magnetic field. 
All interactions that particles might undergo during 
transport, such as energy losses, and \gray{} and synchrotron production 
mechanisms, are similarly included.

%The study of the DGE is one of the priority goals for the 
%forthcoming GLAST mission.
Study of the DGE will advance greatly with the forthcoming GLAST mission.
In the following, we describe our ongoing efforts for understanding and 
modelling the DGE that will be incorporated into the model
for the GLAST Large Area Telescope (LAT) Science Groups.

\section{CR Propagation and GALPROP}

GALPROP is a code for CR propagation and diffuse \gray{} emission.
We give a brief summary of GALPROP; for details we refer to the relevant
papers \cite{Moskalenko1998,Moskalenko2002,Strong1998,Strong2000,Strong2004b}
and a dedicated website\footnote{http://galprop.stanford.edu}.
The propagation equation is solved numerically on a spatial grid,
either in 2D with cylindrical symmetry in the Galaxy or in full 3D.
The boundaries of the model in radius and height, and the grid
spacing, are user-definable. 
Parameters for all
processes in the propagation equation can be specified.  
The
distribution of CR sources can be freely chosen, typically to
represent supernova remnants.  
Source spectral shape and isotopic composition
(relative to protons) are input parameters.
Cross-sections are based on extensive compilations and
parameterisations \cite{Mashnik2004}.
The numerical solution is evolved forward in time until a steady-state is 
reached; 
a time-dependent solution is also an option. 
Starting with
the heaviest primary nucleus considered (e.g., $^{64}$Ni) the
propagation solution is used to compute the source term for its
spallation products, which are then propagated in turn, and so on down
to protons, secondary electrons and positrons, and antiprotons.  
In
this way secondaries, tertiaries, etc., are included.  
Primary electrons are treated separately.  
%The local proton, helium, and electron spectra are 
%normalised to data; all other isotopes are determined
%by the source composition and propagation.
Gamma rays and synchrotron emission are computed  using interstellar gas data 
(for pion-decay and bremsstrahlung) and the ISRF model (for IC).

\begin{figure*}
\begin{center}
\noindent 
%\fbox{\hbox{\vbox{\hsize=130mm \hfill \vspace{50mm}}}}
\includegraphics[height=.30\textheight]{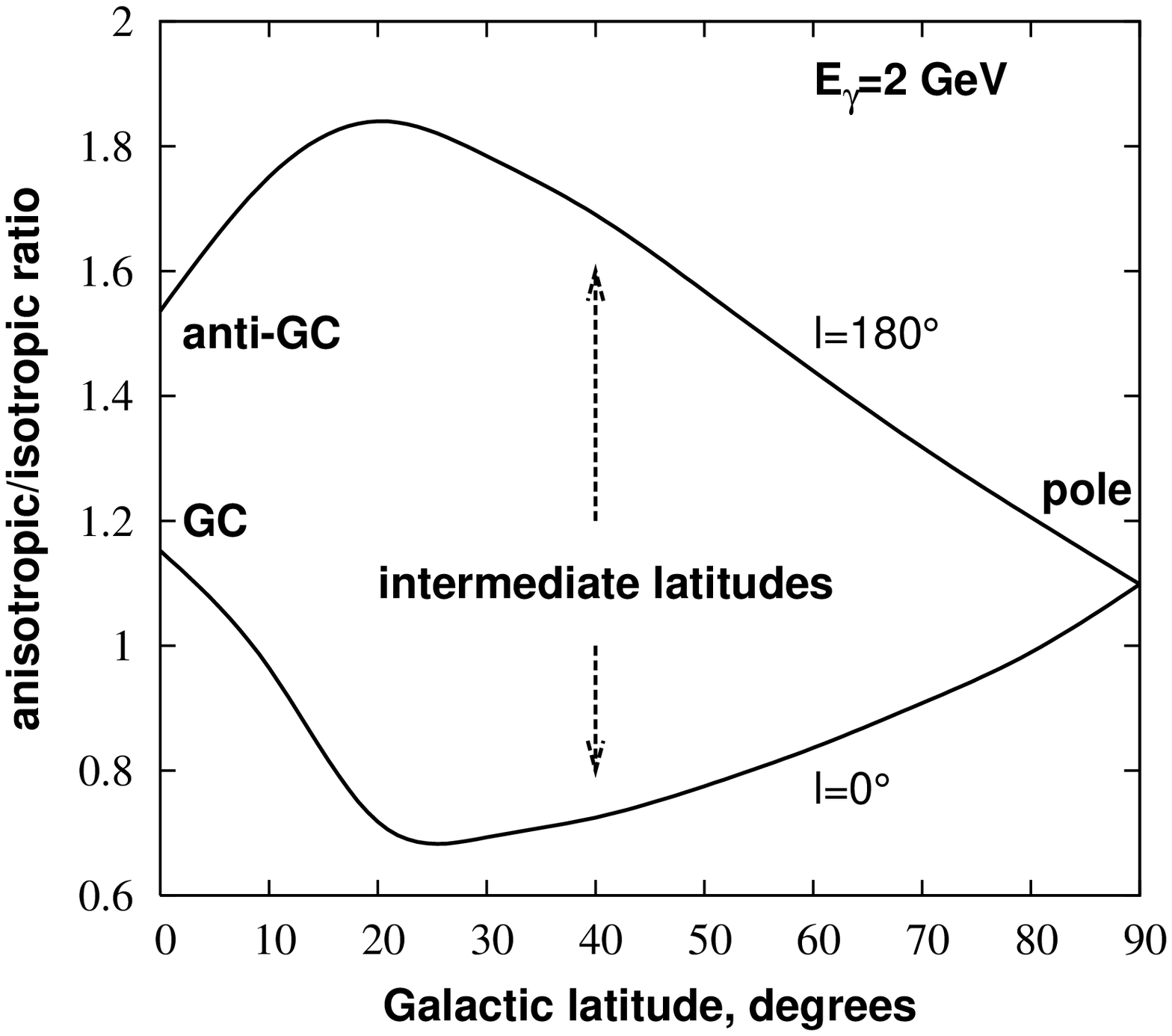}
\includegraphics[height=.30\textheight]{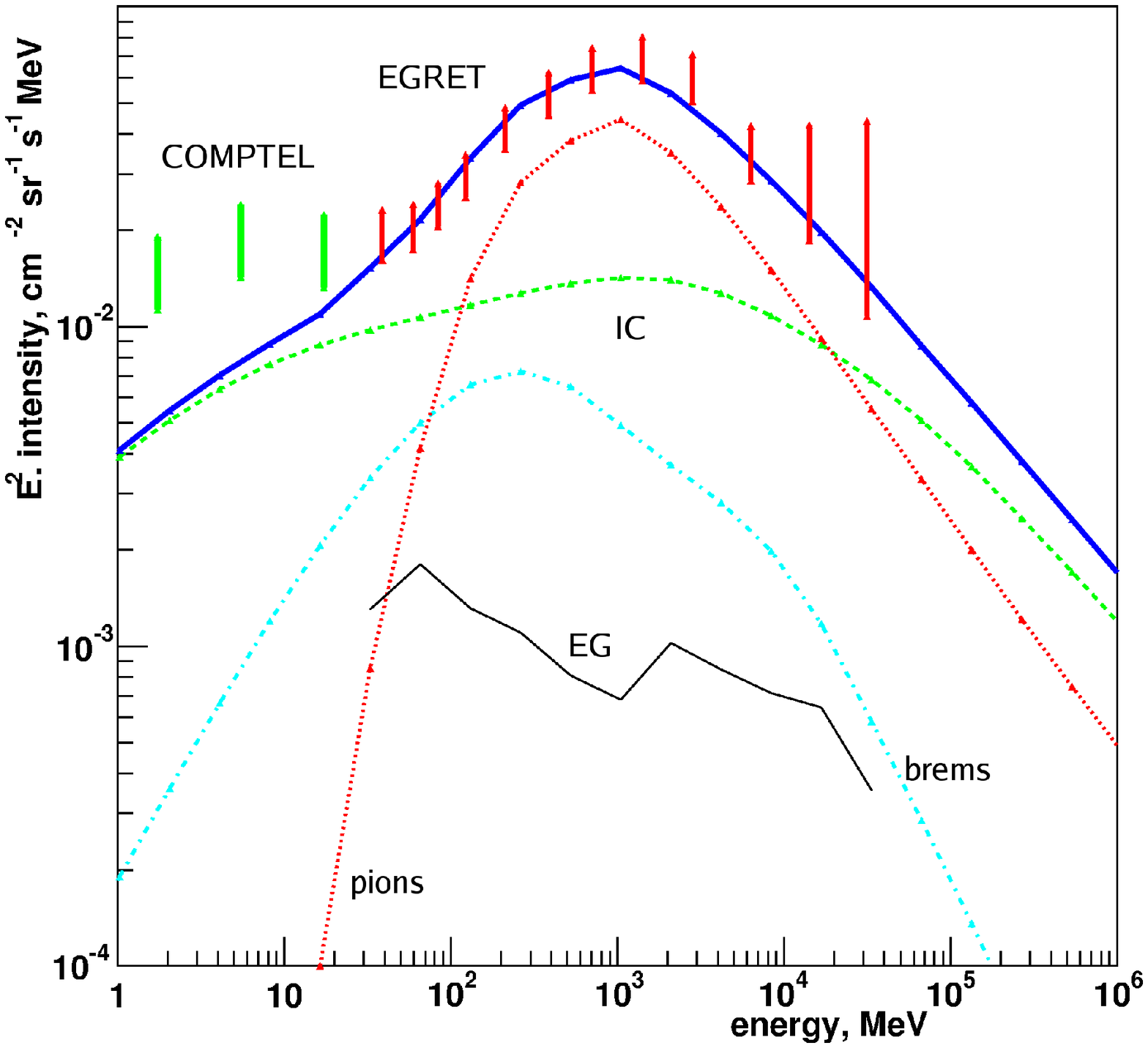}
\end{center}
\caption{{\it Left:} The ratio of anisotropic IC to isotropic IC for Galactic
longitudes $l=0^\circ$ and 180$^\circ$ vs.\ Galactic latitude 
\cite{Moskalenko2007}.
{\it Right:} \gray{} spectrum of inner Galaxy 
($330^\circ<l<30^\circ, |b|<5^\circ$)
for the optimised model. 
Vertical bars: COMPTEL and EGRET data,
heavy solid line: total calculated flux.
This is an update of the spectrum shown in \cite{Strong2004b}.}\label{fig1}
\end{figure*}

We are continuously improving the GALPROP code
to keep up with new theory and data.
%information.
Recent extensions to GALPROP relevant to the GLAST-LAT 
diffuse emission model include
%\begin{itemize}

$\bullet$ 
interstellar gas distributions based on current H~I and CO surveys
(see below) 

$\bullet$
H$_2$ mass calibration ($X_{\rm CO}$-factors) which can vary with Galactocentric
distance

$\bullet$
new detailed calculation of the ISRF (see below)

$\bullet$
proper implementation of the anisotropic IC scattering \cite{Moskalenko2000}
using the new ISRF (Figure~\ref{fig1} [left])

$\bullet$
new parameterisation of the $\pi^0$ production in $pp$-collisions
\cite{Kamae2006} which includes diffractive dissociation

$\bullet$
the extension of the \gray{} calculations from keV to tens of TeV, 
and the production of full sky maps as a function of energy; 
the output is in FITS format (Figure~\ref{fig1} [right])

$\bullet$
a dark matter package to allow for propagation of the WIMP annihilation 
products and calculation of the corresponding synchrotron and \gray{} skymaps

\section{Interstellar Gas}
The maps of the neutral interstellar medium (ISM) used in the 
\gray{} intensity calculations have been updated recently.  
The neutral gas is traced by observations of the 21-cm line of H~I and 
the 115 GHz line of CO (the standard surrogate for H$_2$, which is not directly 
detectable at interstellar conditions).  
The differential rotation of the Milky Way causes distance-dependent 
Doppler shifts of the line frequencies.  
These shifts can be used to derive approximate Galactocentric distances 
for the emitting regions corresponding to the observed spectral lines.  
We use the rotation curve of Clemens \cite{Clemens85} in deriving 
Galactocentric distances and divide the Milky Way into equidistant rings
of $\sim$2 kpc width.  
Because for longitude ranges within 10$^\circ$ of the Galactic centre and 
anticentre so-called kinematic distances cannot be determined, we 
interpolate the maps across these ranges, using a method that ensures 
that the integrated column densities in the interpolated regions are 
consistent with the survey observations.

The new LAB survey of H~I \cite{Kalbera2005} is now used for calculating 
the `rings' of H~I.  
This survey has uniform coverage of the entire sky and has been 
carefully corrected for the effects of stray radiation.  
The CfA composite CO survey \cite{Dame2001} is now used for 
calculating the CO rings.  
The data for the ring that contains the solar circle are augmented 
using a new intermediate and high-latitude survey; 
the initial results of this survey have been published \cite{Dame2004}, but 
additional observations have been made \cite{Dame2007}.

\section{Interstellar Radiation Field}
The large-scale 
ISRF of the Galaxy
is the result of stellar emission and dust reprocessing of the 
star light in the ISM.
There is also a contribution by the cosmic microwave background (CMB).
A model has been constructed for the Milky Way ISRF
\cite{Moskalenko2006a,Porter2005,Porter2007}
incorporating details of stellar population distributions based on 
recent data from surveys such as 2MASS and SDSS, and a radiative 
transfer treatment of dust scattering, absorption, and 
re-emission of the star light in the infrared. 
The dust distribution in the model follows the gas distribution; to ensure
the ISRF is consistent with the GALPROP code we use the gas distributions
described above.

\begin{figure*}
\begin{center}
\noindent 
%\fbox{\hbox{\vbox{\hsize=130mm \hfill \vspace{50mm}}}}
\includegraphics[height=.26\textheight]{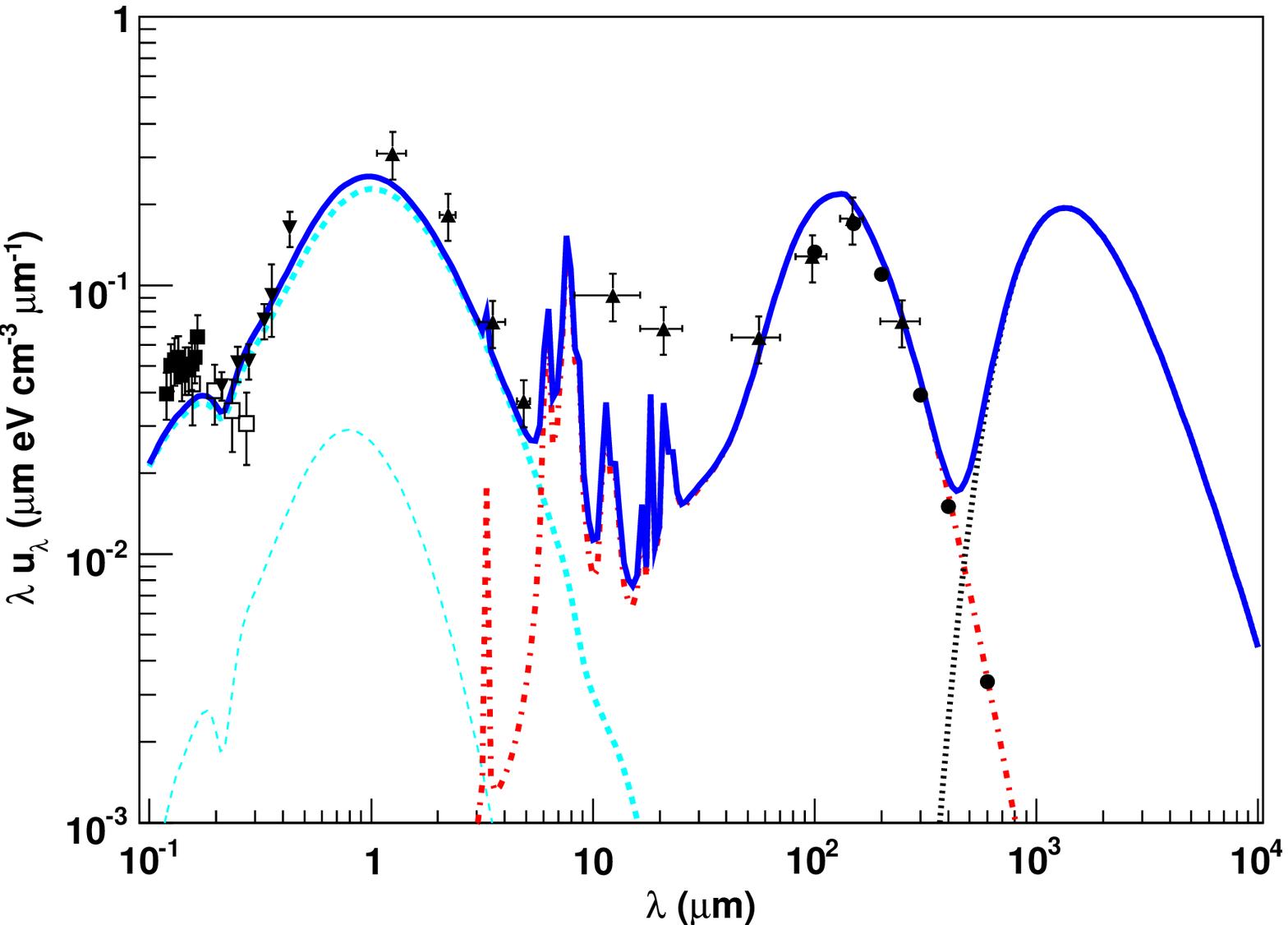}
\includegraphics[height=.26\textheight]{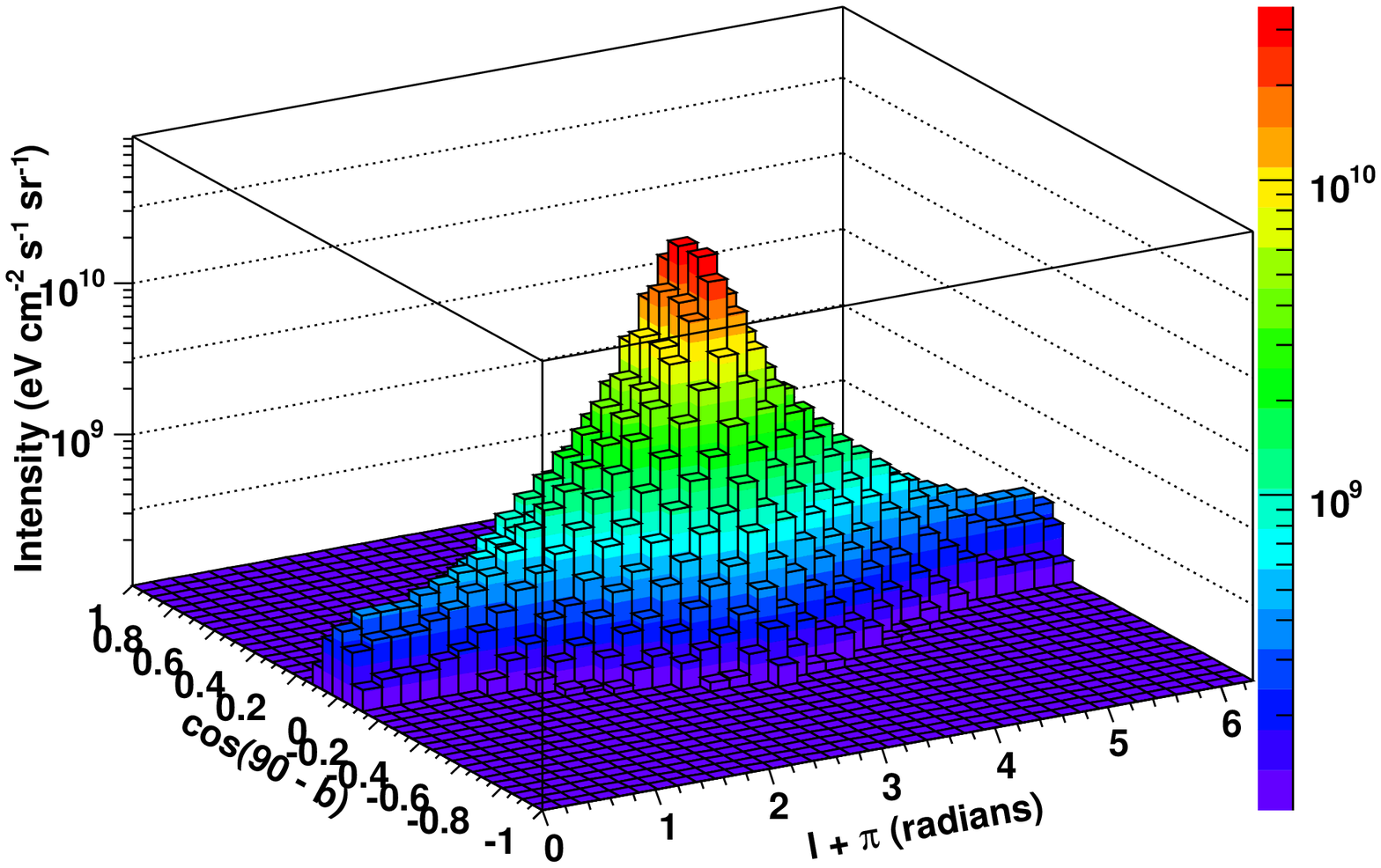}
\end{center}
\caption{
{\it Left:} Local ISRF spectral energy density. 
Line-styles: solid, total; thick dashed, stellar; thin dashed, scattered; 
chain, dust; dotted, CMB. 
Data points are summarised in \cite{Porter2007}.
{\it Right:} 
Model local ISRF intensity at 2.2 $\mu$m as a function of Galactic longitude $l$ and latitude $b$.}\label{fig2}
\end{figure*}

The ISRF model allows the calculation of the spectral energy density (SED) and 
angular distribution as a function of position and wavelength throughout the 
Galaxy.
As an example of the model output, we show in Figure~\ref{fig2} 
the local SED (left) and the local intensity distribution at 2.2 $\mu$m (right).
The SED is important for the CR electron energy losses during propagation. 
The intensity distribution of the ISRF, which was previously not 
available in the literature, allows the calculation of the IC emission 
using the anisotropic IC cross section \cite{Moskalenko2000}.
This has been shown to produce significant differences over the 
sky when compared to the assumption of an isotropic ISRF 
(Figure~\ref{fig1} [left]); the latter approximation is true only for 
the CMB.

\section{Summary}

From the EGRET era we have learned a great deal about the DGE 
in the MeV-GeV range (see, e.g., \cite{Moskalenko2005} and references therein) 
while the GeV-TeV range remains largely unexplored. Recent VHE observations 
\cite{Abdo2007,Aharonian2006} indicate the Galaxy is full of surprises. 
The DGE is also present at multi-TeV energies, but 
large variations
are to be expected because of the inhomogeneity of the sources, and hence CR
distribution; this contrasts with the case in the MeV-GeV range where the 
DGE has a significantly smoother distribution.

We have given a brief summary of work that is being done on the GLAST-LAT 
DGE model prior to launch; naturally, adjustments will be 
required after launch.
The GLAST-LAT will study the DGE in the GeV-TeV range, providing a 
clearer picture and connection between the spacecraft-borne 
and ground-based instruments.
%EGRET regime, and that of VHE instruments.
This will provide 
much new information on CR propagation and sources, and the ISM.
%The launch of GLAST in winter 2008 is eagerly awaited.

\section{Acknowledgements}
%\\
%\\
%{\bf Acknowledgements.}
I.\ V.\ M.\ acknowledges partial support from a NASA APRA grant.
T.\ A.\ P.\ acknowledges partial support from the US Department of Energy.

\end{document}